\documentstyle[preprint,aps]{revtex}
\tightenlines
\begin{document}
\preprint{UM-P-99/51, RCHEP-99/21}
\draft
\title{Geophysical constraints on mirror matter within 
the Earth} 
\author{A.Yu.Ignatiev\cite{byline1} and
R.R.Volkas\cite{byline2}}
\address{School of Physics, Research Centre for High Energy 
Physics, University of Melbourne, Victoria 3010,
Australia}
\maketitle
\begin{abstract}
We have performed a detailed investigation of geophysical constraints
on the possible admixture of mirror matter inside the Earth. On the basis
of the Preliminary Reference Earth Model (PREM)---the ``Standard Model''
of the Earth's interior---we have developed a method which allows 
one to compute changes in various quantities characterising the Earth 
(mass, moment of inertia, normal mode frequencies etc.)due to the presence
 of mirror matter. As a result we 
have been able to obtain for the first time the direct upper bounds 
on the possible concentration of the mirror matter in the Earth.
In terms of the ratio of the mirror mass to the Earth mass a conservative
upper bound is $3.8\times 10^{-3}$. We then analysed possible mechanisms
(such as lunar and solar tidal forces, meteorite impacts and earthquakes)
of exciting mirror matter oscillations around the Earth centre.
Such oscillations could manifest themselves through global variations
of the gravitational acceleration at the Earth's surface. We conclude 
that such variations are too small to be observed. Our results are
valid for other types of hypothetical matter coupled to ordinary 
matter by gravitation only (e.g. the shadow matter of superstring
theories).
\end{abstract}
\pacs{95.35.+d, 91.35.-x, 11.30.Er, 12.60.-i, 14.80.-j}

\section{Introduction}
Recently there has been considerable interest in the study of particle
physics implications of the SuperKamiokande data on atmospheric
neutrinos \cite{v1} as well as the results of other neutrino 
experiments \cite{v2}. The
main purpose of such investigations has been  to understand what
theories could be responsible for the observed experimental features
which give strong evidence for large angle neutrino oscillations. One
of these theories is the Exact Parity Model (EPM) which introduces 
parity or ``mirror'' partners for all ordinary particles
(except the graviton) and thus
restores the parity invariance apparently broken by weak 
interactions \cite{v3,v4,v5,v6,v7}. The Exact
Parity Model predicts pairwise maximal mixing between ordinary
and mirror neutrinos and provides a basis for interpretation of
atmospheric neutrino and solar neutrino data \cite{v5}.

An important question that arises naturally is whether or not the
existence of mirror particles can lead to other observable
consequences. In this work our objective is to find constraints on 
the
possible concentration of mirror particles in the Earth.  These
constraints are a necessary step in the search for mirror world
manifestations  in terrestrial experiments. In particular, the
presence of mirror matter in the Earth could lead to the regeneration
of ordinary neutrinos from mirror neutrinos and consequently to the
suppression of the day-night effect in the neutrino data.

Two main approaches to our problem are possible. First, one can trace
the fate of the mirror particles starting from the early Universe
epoch through the structure formation periods (galaxies, solar
system and finally the Earth). Second, we can use geophysical data to
get a more direct limit on the concentration of mirror matter in the
Earth regardless of possible cosmological bounds.

It has been suggested that considerations based on the 
structure
formation theory disfavour a significant presence of mirror matter in
the Earth (see Blinnikov and Khlopov in \cite{v6}; Kolb, Seckel and 
Turner in \cite{v6}). 
However, as our knowledge of the structure formation is
still incomplete, it is important to develop a geophysical approach 
as
an independent, complementary tool of analysis exploiting the wide and
rich variety of observational data accumulated in the Earth sciences.

This approach will be applicable not only to the specific EPM model,
but also to any other theory predicting the existence of a new world
of particles which couples to the ordinary matter only through
gravitational interaction. An example is the shadow matter 
characteristic
of superstring theories.

The plan of the paper is as follows. Section 2 summarises the main
aspects of the ``Standard Earth Model'' called ``Preliminary 
Reference
Earth Model'' (PREM). In Section 3 we analyse the possible effects of
mirror matter within the context of PREM. Section 4 is devoted to the
study of dynamical manifestations of mirror matter and comparison 
with
gravimetric data. Finally, we present our conclusions in Section 5.

\section{Preliminary Reference Earth Model (PREM)}

The Preliminary Reference Earth Model (PREM) \cite{5} is a 
mechanical model of
the average internal structure of the Earth based, mainly, on the
analysis of seismological data. It gives the radial distributions of
mechanical properties (such as density, elastic moduli, pressure,
gravity and others) in the Earth's interior.

The set of initial data used for constraining the model includes:

1. astronomic-geodetic data (radius, mass and moment of inertia of the
Earth);

2. data on free oscillations and long-period surface waves (over 1000
eigenfrequencies are known);

3. body waves data ( $\sim 10^6$ arrival times have been registered).

Let us summarise the main analytical relations used in the
construction of the model (for more details see e.g. \cite{6,7}).
The velocities of the elastic waves are
given by
\begin{equation}
\label{1}
v_p(r) = \sqrt{{K(r)+{4 \over 3}\mu(r) \over \rho(r)}},\quad v_s(r)=
\sqrt{{\mu(r) \over \rho(r)}},
\end{equation}
where $v_p$ is the velocity of longitudinal waves, $v_s$ is the
velocity of transverse waves, $K$ is the bulk modulus (or 
incompressibility) and $\mu$ is the
shear modulus. These velocities can be found from seismological 
observations as functions of radius $r$. Measurements of wave 
velocities supply the ratios $K/\rho$ and $\mu/\rho$. To obtain 
$\rho$ independently, the Adams-Williamson
equation,
\begin{equation}
\label{2}
-{d\rho \over dr} = \frac{G\rho (r)}{r^2 (v_p^2(r) - 
{4 \over 3}v_s^2(r))}
\int^r_0 4\pi a^2 \rho(a)da,
\end{equation}
must be used. This equation
expresses the condition of mechanical equilibrium between the
gravitational attraction and the pressure due to elastic compression.
$G$ is Newton's constant, and the combination of the squared sound 
speeds
in the denominator is called the seismic parameter $\Phi$:
\begin{equation}
\label{20.1}
\Phi (r)\equiv v_p^2(r) -{4\over 3}v_s^2(r)={K(r) \over\rho (r)}.
\end{equation}
Equation (\ref{2}) is valid for a chemically homogeneous layer with
adiabatic temperature gradient.

Further, the connections between the density profile and the profiles 
of pressure and gravity are given by
\begin{equation}
\label{20.2}
{dP\over dr}=\Phi (r){d\rho\over dr},
\end{equation}
and
\begin{equation}
\label{20.3}
g(r)={G\over r^2}\int^r_0 4\pi a^2\rho(a)da.
\end{equation}
\section{Static constraints}
In this section, we will compute
constraints on {\em stationary} mirror matter from PREM.
A later section will consider dynamic manifestations of 
mirror matter.

\subsection{Pedagogical warm-up exercise}
In order to get a feeling for what would go wrong if a
substantial amount of mirror matter were present in the Earth, we
consider a simple but unrealistic scenario first. Suppose the Earth
is actually two concentric spheres, one ordinary and one mirror.
Suppose also,
for simplicity, that the mass density ratio $\rho_1(r)/\rho_0(r)$
of mirror to ordinary matter is independent of radius.
This is clearly unrealistic and we will relax this assumption
in the next subsection.

In this case, alterations due to mirror matter of the PREM equations 
(\ref{1}, \ref{2}, \ref{20.1}, \ref{20.2}, \ref{20.3}) can be 
accomodated either by rescaling Newton's constant,
\begin{equation}
G \rightarrow G' = G(1 + {\rho_1 \over \rho_0})
\end{equation}
(throughout the paper quantities with index 0 will refer to ordinary
matter while those with index 1 to mirror matter) or by keeping
$G$ fixed and rescaling all of the other quantities used in PREM.

While the first procedure (rescaling $G$) seems to be the simplest, 
the second procedure reveals the physical modifications required
in the presence of mirror matter more clearly and allows one to use
PREM results directly.

From the formulas (\ref{1}) and the Adams-Williamson equation 
(\ref{2}) we deduce that the effective parameters can be defined as
\begin{equation}
\rho_{eff}=\rho_0 + \rho_1,
\end{equation}
\begin{equation}
P_{eff}=P_0(1+\frac{\rho_1}{\rho_0}),
\end{equation}
\begin{equation}
\mu_{eff}=\mu_0(1+\frac{\rho_1}{\rho_0}),
\end{equation}
\begin{equation}
K_{eff} = K_0(1+\frac{\rho_1}{\rho_0}),
\end{equation}
\begin{equation}
g_{eff}=\frac{G}{r^2} \int^r_0 4\pi a^2 \rho_{eff} (a)da,
\end{equation}
\begin{equation}
\Phi_{eff}\equiv v^2_{p,eff} - {4 \over 3}v^2_{s,eff} = {K_{eff} \over
\rho_{eff}}={K_0 \over \rho_0} = v^2_{p,0}-{4 \over 3}v^2_{s,0}
\equiv\Phi_0,
\end{equation}
\begin{equation}
v_{p,eff}=v_{p,0}=\sqrt{{K_{eff}(r)+{4 \over 3}\mu_{eff}(r) \over
\rho_{eff}(r)}},
\end{equation}
\begin{equation}
v_{s,eff}=v_{s,0}=\sqrt{{\mu_{eff}(r) \over \rho_{eff}(r)}}.
\end{equation}
With these definitions the form of the Adams-Williamson equation does
not change:
\begin{equation}
-{d\rho_{eff} \over dr}={\rho_{eff} g_{eff} \over \Phi_{eff}}.
\end{equation}
Also, the usual equation for the bulk modulus holds true in terms of
the effective quantities:
\begin{equation}
{d\rho_{eff}\over dP_{eff}}={\rho_{eff}\over K_{eff}}.
\end{equation}
This demonstrates self-consistency of the rescaling procedure.

Now, for illustrative purposes, consider the case of a 50\%--50\% 
mixture of ordinary and mirror matter. The effective values for
the density, incompressibility and pressure at the Earth centre can be
taken directly from PREM data:
\begin{equation}
\rho_{eff}=13.1\;g/cm^3, \quad K_{eff}=14.2\;Mb, \quad P_{eff}=3.64\;Mb.
\end{equation}
Correspondingly, the values for the ordinary matter (at the
centre) are obtained by a factor of 2 rescaling:
\begin{equation}
\label{8}
\rho=6.55\;g/cm^3, \quad K=7.1\;Mb, \quad P=1.82\;Mb.
\end{equation}

With these values, iron is definitely ruled out as the main component
of the core and, therefore, the Earth as a whole. However, from
independent evidence we know that iron is one the most abundant
elements in the Earth. In addition, no other element (with significant
abundance) can have the properties required by Eq. (\ref{8}). For
instance, silicon (at zero pressure)  has a lower density
than iron, but its incompessibility is {\em greater} than that of
iron. Therefore the 50\%--50\% mixture of ordinary and mirror matter in
the Earth is incompatible with the observational data encoded by the
PREM model taking into account knowledge of terrestrial chemistry.

\subsection{Modified Adams-Williamson equation}
Consider now the realistic case where the mirror matter density does not 
follow
the ordinary density.  Then the density of the ordinary matter
(indexed by 0) would obey the modified Adams-Williamson equation:
\begin{equation}
\label{9}
-{d\rho_0 \over dr} = \frac{G\rho_0}{r^2 (v_{p0}^2 - {4 \over
3}v_{s0}^2)} \int^r_0 4\pi a^2 (\rho_0(a)+ \rho_1(a))da.
\end{equation}
Also, we have to require that the total mass of the Earth and the
moment of inertia are equal to their observed values:
\begin{equation}
\int^{R_{\oplus}}_0 4\pi a^2 (\rho_0(a)+ \rho_1(a))da =M_{\oplus},
\end{equation}
\begin{equation}
\frac{8\pi}{3 M_{\oplus} R_{\oplus}^2} \int^{R_{\oplus}}_0 a^4
(\rho_0(a)+ \rho_1(a))da = I,
\end{equation}
where
\begin{equation}
M_{\oplus} = 5.974 \times 10^{24}\;kg, \quad I=0.3308.
\end{equation}

Let us start with the simplest case: assume that
the mirror matter forms a ball of uniform density $\rho_1$ with a
radius equal to the radius of the inner core ($R_{inner\; core}=1221
km\approx R_{\oplus}/5$).
Within PREM the density profile in the inner core is given by the
following function:
\begin{equation}
\rho_{PREM}(r)=\rho_{PREM}(0) -q{r^2 \over R_{\oplus}^2},
\end{equation}
where
\begin{equation}
\rho_{PREM}(0) = 13.09\;g/cm^3, \quad q=8.84\;g/cm^3.
\end{equation}
Therefore, we look for the solution of the modified Adams-Williamson
equation in the following form 
\begin{equation}
\label{15}
\rho_0(r) = \rho_{PREM}(0) -\epsilon - (q + \delta){r^2 
\over R_{\oplus}^2},
\end{equation}
assuming that the mirror density
$\rho_1$ is small compared to the Earth central density in PREM model
$\rho_{PREM}(0)$.
Solving the system of equations (\ref{9}, \ref{15}) we find:
\begin{equation}
\label{16}
\epsilon \approx 0.84 \rho_1, \quad \delta \approx 0.9 \rho_1.
\end{equation}
Thus we see that if we wish to add to the Earth some amount of mirror
matter then consistency with the equilibrium equation requires that
the density of ordinary matter be decreased (as compared with PREM
density) by approximately the same amount. Also, it can be shown that
if Eq. (\ref{16}) holds then the constraints due to the Earth mass and
the moment of inertia are automatically satisfied as long as 
$\rho_1 < 0.77\;g/cm^3$.

Such a decrease of ordinary matter density can be achieved in one of
two ways:

1) by changing the chemical composition of the core so that the new
composition has lower density than the standard;

2) by decreasing the pressure so that the density of the {\em
standard} core composition is lowered.

Before we consider these two possibilities let us review briefly the
subject of the standard core composition (see e.g. \cite{7}). 
Information about the
chemical composition of the core is obtained by comparing the
mechanical characteristics given by the PREM model with the properties
of various substances under high pressure. In this way it has been
established that the core characteristics are close but not
equal to those of iron. There is sufficient evidence to conclude that
some lighter element should be added to iron in order to satisfy the
geophysical constraint on the core composition. Sulfur and oxygen
appear to be the strongest candidates for that role although other
elements (such as carbon, nitrogen etc) cannot be ruled out at
present.  As examples, cores containing 6--12\% sulfur or 7--8\% oxygen
(by mass) have been proposed as possible compositions consistent with 
the PREM model.

In order to accomodate mirror matter according to the first method
above, we should increase the admixture of light elements as compared
with the standard levels. However, such an increase would typically
raise the
sound velocity in the core (for more details see \cite{7}). 
Requiring that the sound velocity is
equal within accuracy of about 0.3\%  
to the observed value (see \cite{5}) 
we can estimate that
\begin{equation}
\label{17}
\epsilon \alt 0.18\;g/cm^3.
\end{equation}
Next, in the second method we have to lower the pressure in order to
obtain lower density of the (ordinary) matter. Lowering the pressure
would lead to a decrease of the sound velocity and from the same
requirement as above we can conclude that
\begin{equation}
\label{18}
\epsilon \alt 0.06\;g/cm^3.
\end{equation}

Then, using Eq.~(\ref{16}, \ref{17}, \ref{18}) we can obtain the
 upper bound on the
mirror density in the inner core
\begin{equation}
\label{18a}
\rho_1\alt {1\over 0.84}max\{0.18,\;0.06\}\;g/cm^3=0.21\;g/cm^3.
\end{equation}
Translated to the upper limit on the ratio of the mirror mass to the
total mass of the Earth, this becomes:
\begin{equation}
\label{18b}
{M_1\over M_{\oplus}} \alt 2.7\times 10^{-4}.
\end{equation}

\subsection{Constraints from free oscillations of the Earth}     
Let us consider now the effect of changing the ordinary density on the 
Earth eigenfrequencies. Using Eq.(A2--A6) of Ref.\cite{5}
and Eq.(41) of Ref.\cite{bg}
it can be shown that the relative
change of period of the spheroidal $_0S_0$ mode as a result of
changing the inner core density by $\epsilon$ would be 
\begin{equation}
{\delta T\over T}\approx -{0.026\epsilon\over\rho_{PREM}(0)}.
\end{equation}
Requiring that this shift of period be less than the
fitting accuracy, 0.05\%, we obtain a bound on the allowed change
of the ordinary matter density in the inner core:
\begin{equation}
\label{27}
{\epsilon\over\rho_{PREM}(0)}\alt 1.9\times 10^{-2},\quad\epsilon\alt
0.26\;\;g/cm^3.
\end{equation}
Using Eq.~(\ref{16}) this bound can be translated into a limit
on the mirror matter density in the inner core:
\begin{equation}
\label{30a}
{\rho_1\over \rho_{PREM}(0)}\alt 2.3\times 10^{-2},\quad \rho_1\alt 0.3\; 
g/cm^3.
\end{equation}
In terms of the total mass of the mirror matter $M_1$ 
we can rewrite Eq. (\ref{30a}) as
\begin{equation}
\label{100}
{M_1\over M_{\oplus}} 
\alt 3.8\times 10^{-4}.
\end{equation}
Comparing Eqs.~(\ref{18b})  and (\ref{100}) we see that the difference
between these upper limits is not very significant, although 
Eq.~(\ref{18b}) is perhaps a less reliable estimate than Eq.~(\ref{100})
because of
the incomplete knowledge of mechanical properties of various materials
at high pressures characteristic for the Earth's centre. For these
reasons we interpret Eq.~(\ref{100}) as our final conservative upper
bound on the mirror matter mass located inside the inner core of the 
Earth.

\subsection{Arbitrary radius of the mirror matter ball}
So far we have considered the case when mirror matter is contained
completely inside the inner core of the Earth. To justify
such an assumption one would need to know the detailed macroscopic
properties of mirror matter (such as equation of state, chemical
composition etc.); then using the condition of mirror matter
equilibrium one could obtain the relation between the mirror matter
and its radius.
If we do not want to rely on such additional information
than we have to regard the radius of the mirror ball $R_1$ as a free 
parameter of our model. Of course the resulting constraints
will be weaker than they could be otherwise.

For simplicity we will restrict ourselves to the two characteristic
values of $R_1$ in addition to the case $R_1=R_{inner\;core}$ considered 
before:  first, $R_1=R_{outer\;core}\approx 0.55R_{\oplus}$, and, second,
$R_1=R_{lower\;mantle}\approx 0.89R_{\oplus}$. The second choice is 
motivated by the fact that for radii larger than $R_{lower\;mantle}$ the 
Adams-Williamson equation is not valid anymore.

Also, we will assume
that the mirror matter has a uniform density $\rho_1$ while the 
density of the ordinary matter differs from its PREM value by
a radius-independent correction $\delta\rho_0$ for $r\leq R_1$ and
coincides with the PREM value for $r>R_1$. Although both these
assumptions are clearly unrealistic, they nevertheless give us
a self-consistent approximation scheme because both $\rho_1$
and $\delta\rho_0$ are small and therefore the effect of their
radial dependence would be a second-order correction. 

The
precise details here depend on the chemical composition, the
equation of state and other thermodynamic parameters for mirror matter. 
For instance, if we are given
the equation of state for the mirror matter then (using some 
additional simplifying assumptions, such as chemical homogeneity,
neglecting temperature etc.) we could find the density profile of
the mirror matter (in particular, its central density)
from the equation of mechanical equilibrium. However, our goal in
this paper was to obtain limits on the mirror matter that would be
independent of such particularities. 

In any case, our method allows one to calculate the upper limit on
the mirror matter mass for an {\em arbitrary} distribution
of mirror matter $\rho_1(r)$.

As before, our constraints on the mass of the mirror matter will
be based on 4 pieces of information:

1)mass of the Earth and its coefficient of inertia;

2)validity of the modified Adams-Williamson equation;

3)periods of Earth's free oscillations;

4)velocities of elastic waves.

Generally speaking, points 1) and 2) tell us that the correction
to the ordinary matter density should be approximately equal to
the density of the mirror matter $\delta\rho_0\simeq\rho_1$. Next,
using that equality we can compute the shift of the period for
the $_0S_0$ mode and then obtain the upper limit on the mirror mass.
From what follows it will be evident that we do not need to find 
an exact relation
between $\rho_1$ and $\delta\rho_0$ as it would not significantly
change the final constraints. Therefore, the validity of the
modified Adams-Williamson equation (MAWE for short)
can be analysed in a simpler manner than that of Sec. B. 

As a criterion of validity of MAWE we can require that the actual
mass of the Earth (that is, the ordinary mass plus the mirror mass)
inside any radius $r\leq R_1$ should be equal, with the accuracy 
of $1\%$ \cite{5}, to the PREM mass of the Earth (within the same radius): 
\begin{equation}
\label{n1}
w\equiv \left|{M(r)\over M(r)_{PREM}}-1\right|\alt 
1\%,
\end{equation}
where
\begin{equation}
\label{}
M(r)=M(r)_{PREM}+\int^r_04\pi a^2(\rho_1-\delta\rho_0)da\simeq
M(r)_{PREM}+ {4\over 3}\pi r^3(\rho_1-\delta\rho_0).
\end{equation}
It is convenient to introduce the dimensionless ratio 
\begin{equation}
\label{}
f={\rho_1-\delta\rho_0\over{\bar\rho_{\oplus}}},
\end{equation}
where ${\bar\rho_{\oplus}}=5.5\;g/cm^3$ is the average density
of the Earth. In terms of this quantity the condition (\ref{n1})
can be rewritten as
\begin{equation}
\label{n2}
w=\left( {r\over R_{\oplus}}\right)^3 \left( {M_{\oplus}\over 
M(r)_{PREM}}\right) \times
f\alt 0.01,
\end{equation}
for $r\leq R_1$.

Next, we require that the total mass of the Earth equal the 
observed value with an accuracy of $1.3\times 10^{-4}$ \cite{6} which
translates into
\begin{equation}
\label{n3}
\left( {R_1\over R_{\oplus}}\right) ^3\times f\alt 1.3\times 10^{-4}.
\end{equation}

Further, we have to demand that the coefficient of inertia of
the Earth equals its PREM value $I_{PREM}=0.3308$ with an 
accuracy of $\Delta I/I\alt 3\times 10^{-4}$. In terms of $f$
this condition reads
\begin{equation}
\label{}
\left( {R_1\over R_{\oplus}}\right) ^3\left( 1-1.5\left( 
{R_1\over R_{\oplus}}\right) ^2\right) 
\times f\alt 3\times 10^{-4}.
\end{equation}
We observe that this condition does not give us an independent
constraint because it is satisfied automatically as long
as inequality (\ref{n3}) is fulfilled.

Finally, we compute the shift of the period for the $_0S_0$ normal
mode of the Earth due to non-zero $\delta\rho_0$ 
using the same method as in Section C. In the case $R_1=R_{outer\; 
core}$ we find:
\begin{equation}
\label{}
{\delta T\over T}\approx -0.136\left( {\rho_1\over
{\bar\rho_{\oplus}}} -f\right) .
\end{equation}
From inequality (\ref{n3}) we conclude that 
\begin{equation}
\label{}
f<8\times 10^{-4}.
\end{equation}
With these values of $f$ the criterion (\ref{n2}) of MAWE validity 
is clearly fulfilled for all radii $r\leq R_1$. Next, requiring that
the shift of period be less than the fitting accuracy (0.05\%), we
obtain the upper limit on the mirror matter density
\begin{equation}
\label{Z1}
\rho_1 \alt 0.025\;g/cm^3.
\end{equation}
This translates into the following bound on the mirror matter mass:
\begin{equation}
\label{F1}
{M_1\over M_{\oplus}}\alt 7.4\times 10^{-4} \quad \quad 
(for \; R_1=0.55R_{\oplus}). 
\end{equation}

Following the same procedure, in the case $R_1=R_{lower\;mantle}$ 
we obtain:
\begin{equation}
\label{}
{\delta T\over T}\approx -0.096\left( {\rho_1\over{\bar\rho_{\oplus}}} 
-f\right).
\end{equation}
From (\ref{n3}) we find an upper bound on $f$:
\begin{equation}
\label{}
f<1.8\times 10^{-4}.
\end{equation}
Again, the condition of MAWE validity, Eq.~(\ref{n2}), is  
satisfied automatically with these $f$. Further, the upper limit
on the mirror matter density $\rho_1$ becomes
\begin{equation}
\label{Z2}
\rho_1\alt 0.03\;g/cm^3.
\end{equation}
Correspondingly, the bound on the mirror matter is 
\begin{equation}
\label{F2}
{M_1\over M_{\oplus}}\alt 3.8\times 10^{-3} \quad \quad 
(for \; R_1=0.89R_{\oplus}). 
\end{equation}

Comparing our bounds (\ref{Z1}) and (\ref{Z2}) with the limit
(\ref{18a}) we see that the latter limit is significantly weaker
than the former two (although the limit (\ref{18a}) has been obtained
for the inner core, it is also valid for the outer core because of
the similarity of their chemical composition; we also would not
expect substantial changes of this limit in the case of the lower
mantle). Therefore we conclude that Eq.~(\ref{100}), 
Eq.~(\ref{F1}), and Eq.~(\ref{F2}) 
represent our final upper bounds on the mirror matter mass.
Being the largest of the three, Eq.~(\ref{F2}) can also be 
considered as 
the most conservative, radius-independent upper bound on the 
mirror matter mass in the Earth.

\section{Dynamical manifestations of mirror matter}
In this section we are going to analyze constraints on mirror
matter that follow from the precision measurements of the Earth's
gravitational field. Such constraints may arise if the mirror
matter, for some reason, shifts away from the centre of the Earth.

The motion of the mirror matter in the Earth would be controlled
mainly by the Earth's gravity field; inside the core, in a first
approximation this
field grows linearly with the radius:
\begin{equation}
\label{2.1}
g\approx kr,\quad k=3.6\times 10^{-6}\; s^{-2}.
\end{equation}
(The net tidal force exerted on the mirror matter
by the Moon and the Sun is negligibly small as will be discussed
later.) Consequently, the period of the mirror matter motion inside
the Earth $T_1$ would be independent of radius and given by
\begin{equation}
\label{2.2}
T_1 = {2\pi \over\sqrt{k}}= 3311\;s\approx 55\; min.
\end{equation}
This period gives us the time scale for the variation $\delta g$
of gravitational acceleration at the Earth's surface caused by 
the possible
mirror matter motion.

To determine the amplitude of the gravity variation suppose that
the amplitude of the mirror matter motion is $h$. Then the amplitude
of the gravity variation is given by
\begin{equation}
\label{2.3}
{\delta g \over g} \simeq {M_1 \over M_{\oplus}}{h\over R_{\oplus}}.
\end{equation}
The above equation holds exactly only for the circle
on the Earth's surface which lies in the plane of the
mirror matter motion. For points outside this circle Eq.(\ref{2.3})
can still be used for order-of-magnitude estimates.
Requiring that the gravity variation should not exceed the
observational limit, $\delta g/g\alt 10^{-9}$ (see e.g. \cite{8}) 
we obtain an upper
bound on the amplitude of mirror matter motion:
\begin{equation}
\label{2.4}
h\alt  1.7\times 
\left(  {3.8\times 10^{-3}M_{\oplus}\over M_1}\right)  \;m.
\end{equation}

What physical factors could lead to the off-centre shift of
the mirror matter? Let us start by discussing the possible effect
of Moon's gravity. The two key quantities to be considered are the
tidal torque and the net tidal force exerted on the mirror
matter by the Moon. The effect of the tidal torque would be
to slow down the spinning of the mirror matter, in analogy
with the ordinary tidal torque that brakes the Earth's rotation.
The details of the effect depend on the poorly known characteristics
of the mirror matter such as its angular velocity, elastic and
dissipative properties etc.; we will not dwell on these.

On the other hand, the net tidal force and the corresponding
off-centre shift can be estimated without the knowledge of
mirror matter properties. First of all we note that if the Earth
was spherically symmetric then there would be no net tidal force
acting on the mirror ``ball'' placed in the Earth's centre.
However, if the oblateness of the Earth is taken into account
then the net force at the centre is non-zero and thus the centre
of the Earth is not an equilibrium position anymore. The new
equilibrium position for the mirror matter can be found from
the condition of balance between the net tidal force and the
gravitational attraction of the mirror matter by the Earth.

To find the new equilibrium position it is convenient first to
find the point in the Earth where the lunar tidal force vanishes
(``the tidal centre''). 
Due to oblateness of the Earth the positions of the
Earth's centre of mass and the tidal centre are shifted 
relative to each other by a short distance
$b$ (hereafter we ignore the
$18^{\circ}$ inclination of the Moon's orbit relative to the Earth's
equatorial
plane):
\begin{equation}
\label{2.5}
b= J_2 {3R_{\oplus}^2 \over 2R}\approx 171\;m,
\end{equation}
where
\begin{equation}
\label{2.6}
J_2= { C-B \over M_{\oplus}R_{\oplus}^2}\simeq 1.08 \times 10^{-3}
\end{equation}
is the Earth's dynamical oblateness, $R\approx 3.84\times 10^8\;m$
is the distance between the
Moon and the Earth, $C$ and $B$ are the moments of inertia of the
Earth with respect to the principal axes. Therefore the centre of mass
of the mirror matter will move away from the Earth centre (and also
away from the tidal centre) by the distance
\begin{equation}
\label{2.7}
h\approx {2Gmb \over kR^3},
\end{equation}
where $m$ is the Moon's mass. Inserting Eq.(\ref{2.5}) into 
Eq.(\ref{2.7}) we
obtain
\begin{equation}
\label{2.8}
h\approx {3GmJ_2R^2_{\oplus}\over kR^4}\approx 8.1\times 10^{-6}\;m.
\end{equation}
The off-centre shift of mirror matter would create periodic variations
of the gravity acceleration on the surface of Earth:
\begin{equation}
\label{2.9}
{\delta g \over g}\simeq {h \over R_{\oplus}} {M_1 \over
M_{\oplus}}\approx 1.3\times 10^{-12}{M_1 \over M_{\oplus}},
\end{equation}
which is far beyond the observational limits. Thus we have shown that
the effect of a net tidal force due to the Moon is negligible.
A similar result holds for the solar effect:
\begin{equation}
\label{2.10}
\tilde{h}\approx {3GM_{\odot}J_2R^2_{\oplus}\over k\tilde{R}^4}
\approx 10^{-8}\; m
\end{equation}
where $M_{\odot}\approx 2\times 10^{30}\;kg$ is the solar mass,
$\tilde{R}
\approx 1.5\times 10^{11}\;m$ is the distance between the Sun and
the Earth. As expected, the effect of the net tidal force due
to the Sun is even smaller than the lunar effect.

We now consider non-gravitational interactions that
could possibly cause an off-centre shift of the mirror matter. Let us
start by analysing the possible role of 
meteorites and meteor showers colliding with the Earth. 
Suppose that as a result of such
a collision a momentum $p$ is transferred to the Earth. Then the mirror
matter (assumed to be at rest in the centre) would receive an initial
velocity $u=p/M_{\oplus}$ relative to the Earth. Therefore, the off-
centre displacement would be equal to
\begin{equation}
\label{col}
h_{col}={u\over \sqrt{k}}={p\over M_{\oplus}\sqrt{k}}.
\end{equation}
What could be the magnitude of $p$? The maximal velocity of a
Sun-bound colliding object, relative to the Earth, is 
$v_{max}\simeq 73\;km/s$.
The heaviest meteorite found on the Earth has the mass $m_{max}\simeq
60\;ton$. Inserting these values into Eq. (\ref{col}) we obtain:
\begin{equation}
\label{col1}
h_{col}\alt{m_{max}v_{max}\over M_{\oplus}\sqrt{k}}\approx
3.8\times 10^{-13}\;m.
\end{equation} 
The
variations of surface gravity acceleration caused by such
displacements are many orders of magnitude
beyond observational accuracy. Note that
the impact of meteor showers would be much less than the estimate
(\ref{col1}) since the total mass of even the most copious showers is
significantly less than $m_{max}$.

In the case of still heavier meteorites which disperse after 
hitting the Earth the mass can be estimated only indirectly
(see e.g. \cite{m}).
For instance, the meteorite that created the Arizona crater 
(with diameter of 1207 m and depth 174 m) had the estimated mass
between 60 and 200 thousand tons. The mass of the Tunguska meteorite 
(fell in Siberia in 1908) was at least 1 million tons, 
its speed 30--40 km/s; if we insert 
these values into Eq.~(\ref{col}), we obtain
\begin{equation}
\label{}
h_{col}\simeq 10^{-8} \;m;
\end{equation}
corresponding values of $\delta g/g$ are still completely
negligible (even in comparison with the variation of $g$ due
to the Moon's tidal effect, Eq.~(\ref{2.9})).

Let us now turn to another possible mechanism of the mirror matter
motion. Can earthquakes cause translational oscillations of mirror
matter around the Earth centre? Large enough earthquakes are known 
to excite
free vibrations of the Earth; the study of these vibrations has become
one of the most important pieces of information about the Earth
interior (for more details see e.g. \cite{6,9,F,10}). 
These vibrations are classified into two categories:

---toroidal, in which only shear strain is present so that density is
not perturbed; they are denoted by $_rT_l$;

---spheroidal, where both shear and volume deformations arise, denoted
by $_rS_l$. The indexes $r$ and $l$ (for both $S$ and $T$ modes) have
the same meaning as the radial and orbital quantum numbers of the
hydrogen atom.

Toroidal modes do not lead to gravity perturbations so they cannot
excite oscillations of mirror matter. On the other hand, spheroidal
modes might cause the excitation of mirror matter oscillations through
the gravitational coupling. Note that we should distinguish between
two possible types of mirror matter oscillatons:
a) bulk vibrations in which the centre of mass of mirror matter stays
at rest in the Earth's centre and
b) translational oscillations where the centre of mass of mirror
matter oscillates around the Earth's centre in the Earth's
gravitational field.
We cannot say much about the spectrum of bulk vibrations without
knowing the detailed structure of mirror matter (i.e., its density,
elastic and dissipative properties etc.); for this reason we leave
them out of our consideration. On the contrary, the period of {\em
translational} oscillations can be found and is given by
Eq.(\ref{2.2}).

Our next task is to find out if there are any spheroidal Earth
eigenmodes that could resonate with translational oscillations of
mirror matter. Note that here we deal with the case of {\em
parametric} resonance so we need to look for the eigenperiod
$T_E=T_1/2\approx 1655\;s$ rather than $T_E=T_1$ (the parametric
resonance in the case $T_E=T_1$ is weaker than for $T_E=T_1/2$). The
closest such eigenmode is $_0S_4$ with the period of \cite{5}
\begin{equation}
\label{2.13}
T(_0S_4)=1545.6\; s.
\end{equation}

In the time-varying gravitational field of $_0S_4$ mode the frequency
of translational mirror matter oscillations also becomes time
dependent according to the law
\begin{equation}
\label{2.14}
\omega^2(t)=\omega^2_1(1+a\cos{\gamma t}),
\end{equation}
where
\begin{equation}
\label{2.15}
\omega_1={2\pi\over T_1},\quad \gamma={2\pi\over T(_0S_4)},\quad
a={\delta \rho\over \rho},
\end{equation}
${\delta \rho/\rho}$ is the amplitude of density variation in the
$_0S_4$ mode.

The onset of parametric resonance is controlled by the quantity $s$
called amplification index (see e.g. \cite{10}):
\begin{equation}
\label{2.16}
s={1\over 2}\sqrt{\left( {a \omega_1\over 2}\right) ^2 -\epsilon^2},\quad
\epsilon=\gamma-2 \omega_1.
\end{equation}
If the amplification index is real then the oscillation amplitude
grows with time as $\exp{st}$. In the opposite case $s^2\leq 0$
parametric resonance does not occur.

Using Eqs.~(\ref{2.14},\ref{2.15},\ref{2.16}) we can find $s$ to be
\begin{equation}
\label{2.17}
s={\omega_1\over 4}\sqrt{\left( {\delta \rho\over \rho}\right) ^2 
- (0.28)^2},
\end{equation}
which is clearly imaginary. Thus we conclude that the condition for a
parametric resonance is not satisfied and consequently there is no
amplification of translational oscillations of the mirror matter.

It can be shown that the condition of parametric resonance with the
$_0S_2$-mode takes the following form:
\begin{equation}
\label{2.18}
-{5\over 24}a^2\omega_1<\gamma ' - \omega_1<{1\over 24}a^2\omega_1,
\end{equation}
where
\begin{equation}
\gamma '={2\pi\over 3233.25}\;s^{-1}
\end{equation}
is the frequency of the $_0S_2$-mode.
One can see that the condition (\ref{2.18}) is not satisfied and there
is no resonance with the $_0S_2$-mode either.
\section{Conclusion}
We have investigated in detail geophysical constraints on the possible
admixture of mirror matter inside the Earth. To this purpose, a method
has been developed based on
the Preliminary Reference Earth Model---the ``Standard Model''
of the Earth which describes its internal structure derived from the
geophysical data in a systematic and self-consistent manner. 
If the density of the mirror matter is given, our method allows one to 
compute changes in various quantities characterising the Earth (such as
its mass, moment of inertia, frequencies of its normal modes etc.).
Comparing the computed and observed values of these characteristics,
we can obtain for the first time the direct upper bounds on the
possible concentration of the mirror matter in the Earth. In terms of
the ratio of the mirror mass to the Earth mass these upper bounds range
from $3.8\times 10^{-4}$ to $3.8\times 10^{-3}$ depending on the
radius of the mirror matter ball. 
 We then analyzed possible manifestations
of mirror matter through the variations of the gravity acceleration on
the Earth surface. These variations could arise as a result
of an off-centre shift of the mirror matter due to several possible
mechanisms such as lunar and solar tidal forces, meteorite impacts
 and earthquakes.
Our estimates have shown that variations caused by these mechanisms
are too small to be observed.

In this work we have been based on a standard premise that mirror
matter interacts with ordinary matter only gravitationally
\footnote{Note that mirror matter can also couple to ordinary
matter through photon---mirror-photon mixing \cite{v4,new}. 
An analysis of this
interesting possibility is beyond the scope of the present work.};
we have
not relied on any other specific assumptions about the mirror matter
properties. Therefore our results are valid for other types of
hypothetical matter coupled to ordinary matter by gravitation only; an
example is shadow matter introduced in string theories. On the other
hand, the use of equation of state and other macroscopic characteristic
of mirror matter could lead to more severe constraints on the mirror 
mass inside the Earth.

\acknowledgments
The authors are grateful to G.C.Joshi, R.Foot and B.H.J.McKellar for  
interesting discussions. This work was supported in part by the 
Australian Research Council.

\end{document}